\documentclass[twocolumn, trackchanges]{aastex701} 

\usepackage{graphics,epsf}
\usepackage[utf8]{inputenc}
\usepackage{amsmath}                
\usepackage{amsfonts}               
\usepackage{amssymb}                
\usepackage{epsfig}                 
\usepackage{graphicx}               
\usepackage{float}
\usepackage{color}
\usepackage{multirow}               
\usepackage{hyperref}

\hypersetup{
    colorlinks=true,
    linkcolor=red,   
    urlcolor=cyan}

\usepackage[colorinlistoftodos]{todonotes}

\newcommand{\km}{{~\rm km}}
\newcommand{\s}{{~\rm s}}

\newcommand{\g}{{~\rm g}}
\newcommand{\G}{{~\rm G}}
\newcommand{\K}{{~\rm K}}
\newcommand{\erg}{{~\rm erg}}
\newcommand{\yr}{{~\rm yr}}

\newcommand{\kpc}{{~\rm kpc}}

\newcommand{\keV}{{~\rm keV}}


\begin{document}

\title{The precessing jet axis of the supernova remnant 3C 397}

\author[0009-0009-9708-6915]{Aleksei Klimov}
\affiliation{Department of Physics, Technion - Israel Institute of Technology, Haifa, 3200003, Israel \\ aleksei.k@campus.technion.ac.il; s.dmitry@campus.technion.ac.il; soker@technion.ac.il}
\email{aleksei.k@campus.technion.ac.il}
\correspondingauthor{Aleksei Klimov}

\author[0000-0002-9444-9460]{Dmitry Shishkin}
\affiliation{Department of Physics, Technion - Israel Institute of Technology, Haifa, 3200003, Israel \\  aleksei.k@campus.technion.ac.il; s.dmitry@campus.technion.ac.il; soker@technion.ac.il}
\email{s.dmitry@campus.technion.ac.il}

\author[0000-0003-0375-8987]{Noam Soker}
\affiliation{Department of Physics, Technion - Israel Institute of Technology, Haifa, 3200003, Israel \\ aleksei.k@campus.technion.ac.il; s.dmitry@campus.technion.ac.il; soker@technion.ac.il}
\email{soker@technion.ac.il}

\begin{abstract}
We identify an S-shaped morphological feature in the enigmatic supernova remnant (SNR) 3C 397, which we attribute to the shaping by a precessing pair of jets during the explosion. We identify an S-shaped, faint region composed of two bubbles, located to the north and south of the center, between two X-ray-bright sides.  We attribute the S-shape to a pair of precessing jets that were part of the explosion process. The identification of a main jet axis in SNR 3C 397 increases its similarity to the enigmatic SNR W49B. We discuss two possible scenarios for SNR 3C 397 and W49B. (1) The thermonuclear common-envelope-jet supernova scenario, which was suggested before for W49B, where a neutron star destroys a white dwarf and accretes part of the white dwarf's material via an accretion disk that undergoes a thermonuclear outburst and launches the jets.  (2) The collapse-induced thermonuclear jet-driven explosion, which is a core-collapse supernova driven by jets, as in the majority of, or even all, core-collapse supernovae, and in addition, there is a thermonuclear outburst of a rare helium-oxygen mixed layer in the core, which is triggered by the core collapse. Our study emphasizes the primary role of jets even in the enigmatic SNR 3C 397.  
\end{abstract}


\section{Introduction}
\label{sec:Introduction}

The supernova remnant (SNR) 3C 397 (G41.1-0.3) is an enigmatic remnant on two counts: its morphology is unusual, with a large, rectangular (box-like) shape (e.g., \citealt{DyerReynolds1999, Castellettietal2021}) and a protrusion to the southeast, and its composition does not fit the expectation of a typical core-collapse supernova (CCSN) or a type Ia SN (SN Ia), because it requires a very dense white dwarf (WD) progenitor (e.g., \citealt{Daveetal2017, MartinezRodriguezetal2017, Ohshiroetal2021}). 
It is a bright radio (e.g., \citealt{AndersonRudnick1993, LeahyRanasinghe2016}), infrared (e.g., \citealt{Reachetal2006, Velazquezetal2025}), and X-ray source (e.g., \citealt{SafiHarbetal2000, SafiHarbetal2005, JiangChen2010}), at a distance of $\simeq 8-10 \kpc$ \citep{LeahyRanasinghe2016, MartinezRodriguezetal2020}; \cite{Treyturiketal2026} scale the distance with $8.5 \kpc$. 
\cite{Velazquezetal2025} estimate its age as $\simeq 1000 \yr$, younger than earlier estimates.  Metals are distributed in clumps (e.g., \citealt{Ohshiroetal2021}). 

\cite{ChenYetal1999} and \cite{SafiHarbetal2005} argued, based on their analysis of X-ray emission, that 3C 397 interacts with a denser medium on the western side.  
\cite{Velazquezetal2025} proposed the formation of the rectangular shape as resulting from the expansion of an SN Ia ejecta into a circumstellar medium sculpted by the stellar wind of the SN-progenitor companion star. They simulated this process, including an ambient magnetic field. We agree that an interaction with the ambient medium occurs, but in this study we suggest that the inner structure was shaped by an energetic precessing pair of jets rather than by a dense equatorial gas.  

\cite{Itoetal2025} present CO observations of 3C 397. Their conclusion is that the CO cloud at the local standard of rest systematic velocity of $\simeq 60 \km \s^{-1}$ is circumstellar material ejected by a giant companion to the WD that exploded, i.e., a single-degenerate scenario, and that this cloud interacts with the SNR ejecta. They find spatial correspondence between the radio morphology of SNR 3C 397 and the CO cloud in some regions, but not all. \cite{Itoetal2025} claims that SNR 3C 397 does not interact with the cloud at the local standard of rest systematic velocity of $\simeq 32 \km \s^{-1}$, the cloud which \cite{Jiangetal2010} argued that 3C 397 interacts with. Based on the cloud at  $\simeq 32 \km \s^{-1}$, \cite{Chenetal2013} estimated the progenitor mass to be $\simeq 12 M_\odot$; but in light of the claim of \cite{Itoetal2025} that 3C 397 does not interact with this cloud, this mass estimate is outdated. Most studies classified SNR 3C 397 as resulting from an SN Ia (e.g., \citealt{Yangetal2013}) or a peculiar SN Ia (e.g., \citealt{Mehtaetal2024}), based on its composition.   
\cite{Yamaguchietal2014} and \cite{Siegeletal2021} could not determine the origin of 3C 397 (SN Ia or CCSN) from the Fe K$\alpha$ line energy alone. 

We accept that the interaction of the ejecta with a cloud leads to strong radio emission and some shaping of the outer zones. However, in Section \ref{sec:Sshaped} we will argue that the main shaping of the inner ejecta regions, as revealed by X-ray emission, is due to a precessing pair of jets during the explosion.

\cite{Mehtaetal2024} performed hydrodynamical simulations and conclude that 3C 397 was a pure deflagration explosion. They can reproduce the high mass ratios of Mn/Fe, Ni/Fe, and Cr/Fe in 3C 397, as reported by \cite{Yamaguchietal2015}, who analyzed a \textit{Suzaku} X-ray spectrum. 

Recently, \cite{Treyturiketal2026} studied the two SNRs 3C 397 and W49B, and compared them. Although they find that certain thermonuclear explosion models best match the observed Fe/Si and Ca/Si ratios in both SNRs, no model fully reproduces the complete set of observed abundance ratios, and the explosion should be of low energy, $E_{\rm exp} \simeq 10^{50} \erg$. They also argued that in 3C 397, high Fe enrichment and spatial abundance variations suggest interaction with a dense progenitor environment. 
They claim that W49B's composition is better compatible with a thermonuclear supernova (SN Ia or peculiar SN Ia), but note that the analysis does not account for asymmetry in the ejecta yields and does not satisfactorily match the numerical models they considered. 

This study is motivated by the morphologies of SNRs 3C 397 and W49B. Despite claims that W49B is of thermonuclear origin, we have found a clear indication of shaping by an energetic pair of jets (e.g., \citealt{BearSoker2017, SokerShishkin2025W49b}); such pairs are not expected in SNe Ia. A scenario that includes both jets and a thermonuclear explosion is the one proposed by  \cite{GrichenerSoker2023} for W49B: the thermonuclear common-envelope jet supernova (CEJSN) scenario. In this scenario, a neutron star (NS) destroys the core of a red supergiant at the end of a common envelope evolution and accretes some of the core mass. Some of the core material undergoes a thermonuclear outburst, giving rise to the observed composition. The NS accretes the material via an accretion disk that launches the jets. 
The remnant is either a massive NS or a black hole. A NS remnant is not observed because that NS started as a cool NS, and might have faded within hundreds of years. 
\cite{GrichenerSoker2023} crudely estimated that there is such an event per 200 CCSNe. Motivated by the similarities between SNRs 3C 397 and W49B, we search for jets' signatures in the morphology of 3C 397 (Section \ref{sec:Sshaped}), and further explore its morphology in Section \ref{sec:Perpendicular}. In Section \ref{sec:W49B} we compare similar morphological features in SNRs 3C 397 and W49B. 

\cite{Treyturiketal2026} discussed the early claims that SNR 3C 397 is due to a CCSN because of its unusual morphology, proximity to the Galactic
plane, and interaction with a molecular cloud (e.g., \citealt{SafiHarbetal2000}), and the later claim of a thermonuclear explosion based on the composition (e.g., \citealt{MartinezRodriguezetal2020}). They did not consider the thermonuclear CSEJSN scenario. In our summary section (Section \ref{sec:Summary}), we discuss this possibility, as well as a scenario of a CCSN with a thermonuclear outburst, although the jets supply most of the energy.   

\section{Identifying S-shaped morphology}
\label{sec:Sshaped}

Chandra X-ray telescope observed SNR 3C 397  in 2001 (observation ID 1042; \citealt{SafiHarbetal2005}) with the ACIS-S instrument for a total exposure time of 66 ks. We processed the data using CIAO software (\citealt{Fruscione2006}; Version 4.17), as described in the official Chandra X-ray Center thread. The Images were analyzed using SAOImageDS9 (\citealt{William2017}). To emphasize large-scale features of the SNR, we performed Gaussian smoothing, characterized by the radius $r$ over which we average and the width $\sigma$. This smooths small-scale features that are usually resolved by Chandra. We used different values for these parameters: maximum smoothing $r=20~{\rm pixels}$ and $\sigma=10~{\rm pixels}$; medium smoothing: $r=11~{\rm pixels}$ and $\sigma=5.5~{\rm pixel}$; minimum smoothing: $r=2~{\rm pixels}$ and $\sigma=1~{\rm pixel}$

\begin{figure*}[]
	\begin{center}
\includegraphics[trim=0.0cm 0.0cm 0.0cm 0.0cm ,clip, scale=0.7]{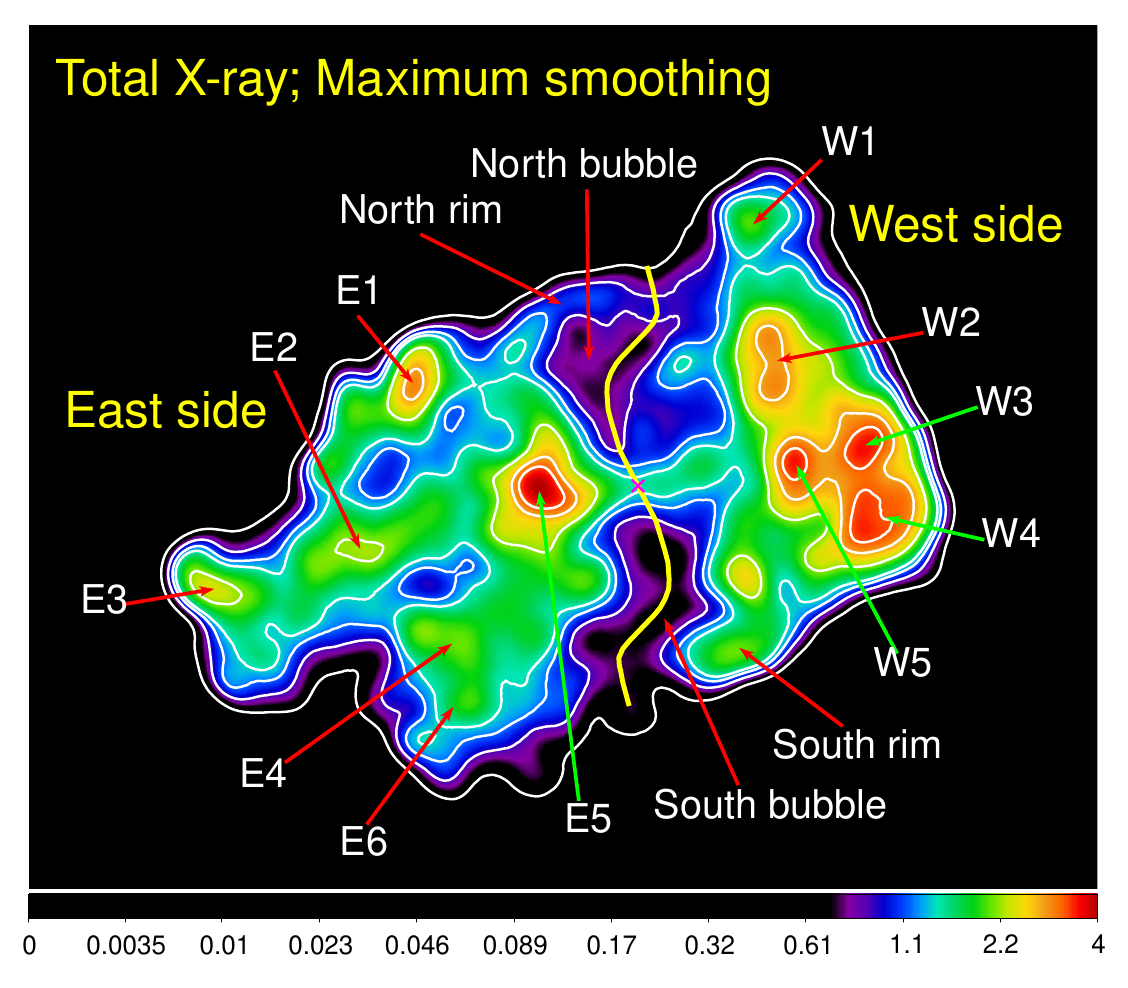} 
\caption{
A Chandra $0.5-8 \keV$ X-ray image of SNR 3C 397. The data were processed using Gaussian smoothing with a radius of $r=20~{\rm pixels}$ and $\sigma=10~{\rm pixels}$. White contours mark the count levels for this smoothing with values of $({\rm count/pixel}) = 0.55$, $0.93$, $1.2$, $1.4$, $2.1$, $2.8$, $3.25$. 
The units of the color-bar are counts/pixel. 
We mark several bright clumps on each side of the remnant, and the faint zones extending north and south of the bright bar near the center, which we term the north and south bubbles. The north and south rims extend from one side towards the bubbles. 
The S-shaped profile we draw with the yellow curve passes through the two bubbles; it is point-symmetric around the `X' point, i.e., the north and south parts are $180^\circ$ apart. We attribute the formation of the two S-shaped bubbles to a precessing jet pair during the explosion.   
}
\label{Fig:MaxSmLarge}
\end{center}
\end{figure*}

To identify the large-scale morphological features, we present in Figure \ref{Fig:MaxSmLarge} the X-ray image of SNR 3C 397 in the entire energy band we consider here, $0.5 - 8 \keV$, and with maximum Gaussian smoothing. We mark several bright and faint regions to aid in exploring the morphology. Earlier studies (e.g., \citealt{SafiHarbetal2005}) already noted the bright west and bright east sides (sometimes called lobes) and the clumpy, irregular morphology of SNR 3C 397. Our new identification is the S-shaped morphology of the faint, north-south-extending central region. We mark an X in red inside the east-west bar. We will refer to this point as the center of the point-symmetric structure we identify.    

There are two bright point-like X-ray sources in the field of view of Chandra, which were previously identified by \cite{SafiHarbetal2005}. The first source at $\alpha = 19^{\mathrm h}07^{\mathrm m}38^{\mathrm s}\!.3, \delta = +07^\circ09'22''\!.9$, just northwest of the peak of E1,  is a soft X-ray source with a column density much lower than that of SNR, ruling out its association with 3C 397. The second source is CXO J190741.2+070650 - a hard X-ray source with prominent iron emission, whose true nature is still unidentified to this day \citep{SafiHarbetal2005, RodriguezToman_2025_proceedingCXO}. It is located outside the SNR to the southeast. Since the origins of those two sources are outside the scope of this paper, we removed them from the images. 

\begin{figure}[h!]
	\begin{center}
\includegraphics[trim=0.0cm 0.0cm 0.0cm 0.0cm ,clip, scale=0.44]{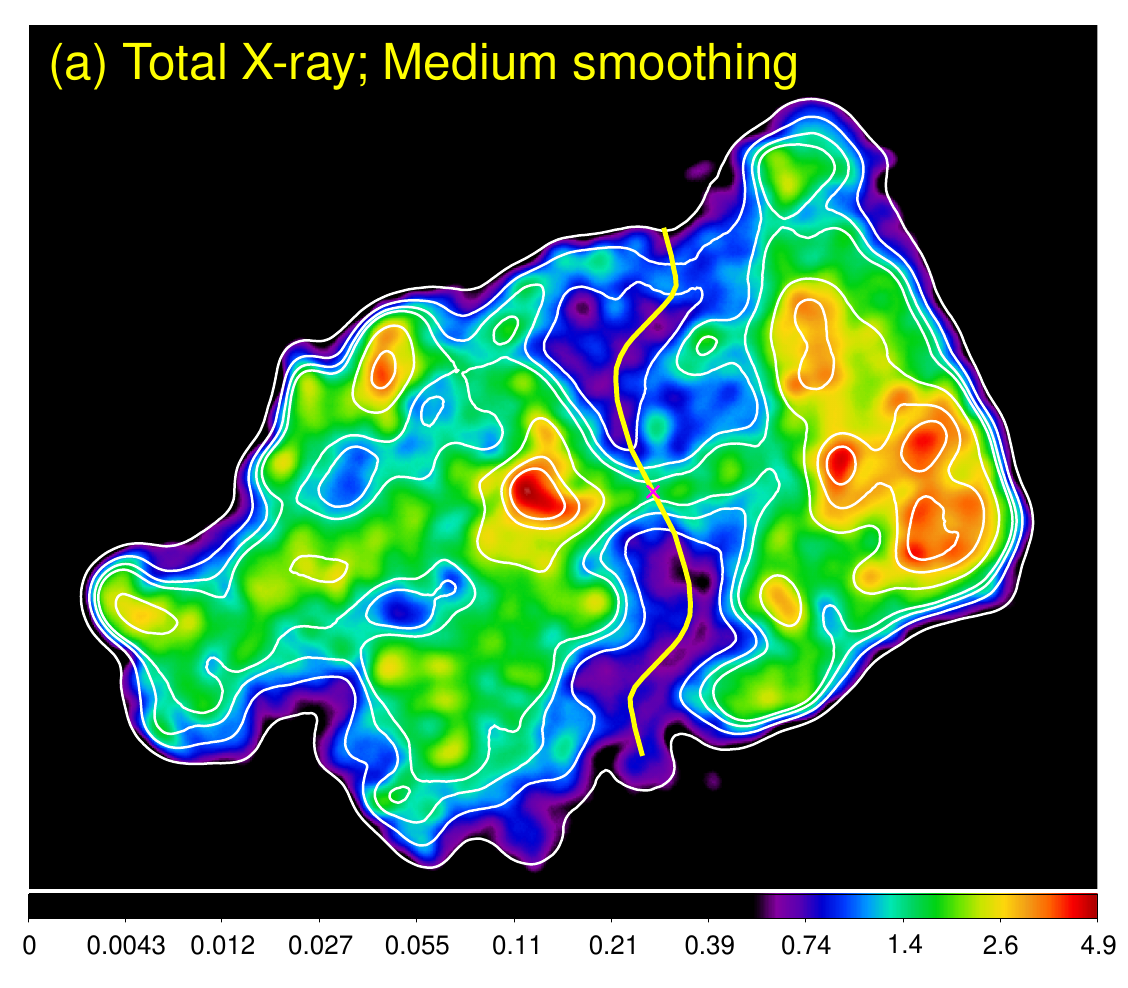}
\vskip 0.05 cm 
\includegraphics[trim=0.0cm 0.0cm 0.0cm 0.0cm ,clip, scale=0.44]{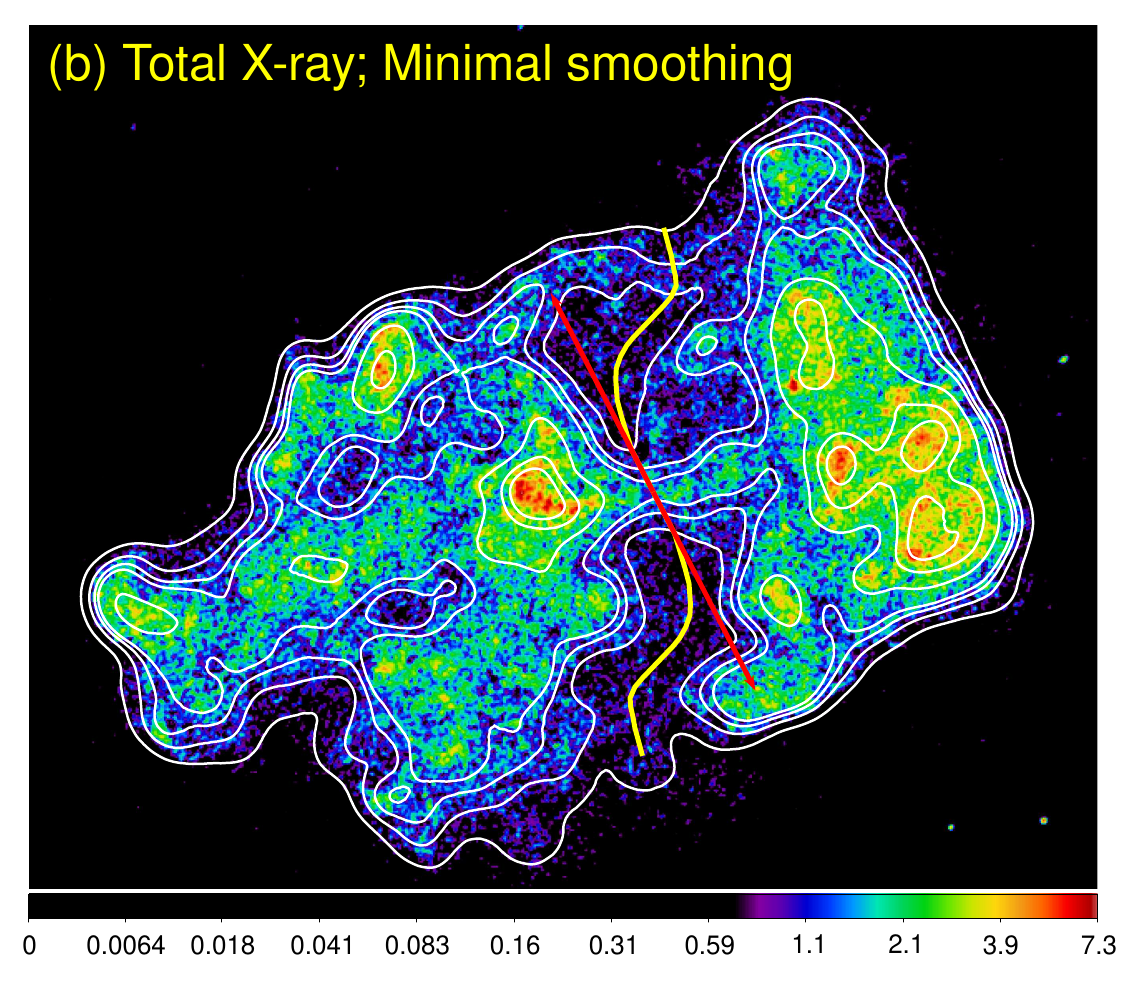}
\caption{
 (a) Same data, contours, and S-shaped yellow line as in Figure \ref{Fig:MaxSmLarge}, but a finer Gaussian smoothing with $r=11~{\rm pixels}$ and $\sigma=5.5~{\rm pixels}$. Values of colors according to the color bar in units of (count/pixel). (b) Same data, contours, and S-shaped yellow line as in panel (a), but an even finer Gaussian smoothing with $r=2~{\rm pixels}$ and $\sigma=1~{\rm pixel}$. 
 Values of colors according to the color bar in units of count/pixel.
 The double-sided red arrow has its center at the `X', i.e., the center of the S-shaped yellow line, and points to the north and south rims to emphasize the point-symmetric structure of the bubbles' boundaries. Note the different color bars in the two panels and from Figure \ref{Fig:MaxSmLarge}.   
}
\label{Fig:MediumSm}
\end{center}
\end{figure}

We notice faint north- and south-facing regions, which we term bubbles, on either side of the central east-west bar. We drew a yellow line through the middle of the south bubble, starting from the center (red-X on the bar). We then copied and rotated it by $180^\circ$ around the center. The copied northern line goes through the faintest parts of the north bubble, and the entire yellow line has an S-shape. We attribute this S-shaped bubble pair to inflation by a pair of precessing, opposite jets. The bubble pair maintains its S-shape in the medium and minimum Gaussian-smoothed images we present in Figure \ref{Fig:MediumSm}. Note that the color in each of the two panels is according to the smoothing of the panel, but the contours are the same as in Figure \ref{Fig:MaxSmLarge}, i.e., corresponding to maximum Gaussian smoothing. The yellow S-shaped line is the same as in Figure \ref{Fig:MaxSmLarge}. To emphasize the point-symmetric structure of the bubbles and their surroundings, we draw a double-sided red arrow on panel (b) of Figure \ref{Fig:MediumSm} with its center on the SNR center (red X), pointing to the south and north rims.  

To visualize the emission of different elements, we prepared images of equivalent widths, following a similar approach to \cite{UnaHwang2000}. We used the \textit{dmimgcalc} command in CIAO to calculate the values for each pixel of the picture separately. We took the emission (which is the sum of continuum and line) of Si in the $1.78 - 1.9 \keV$ band, S in the $2.35 - 2.5 \keV$ band, and Fe in the $6.3 - 6.8 \keV$ band. The continuum is assumed to be linear around the line emissions. The continuum counts under the line emission were estimated as the average of the continuum counts to the left and right of the line. The average continuum was then subtracted from the total counts. We calculated the equivalent width by dividing the line counts by the continuum counts. To increase the signal-to-noise ratio, the data were binned, combining a 4-by-4-pixel area into a single bin for Si and S images and a 16-by-16-pixel area for the Fe image. We chose larger binning for the Fe image to account for the weaker continuum emission in the Fe line, with 16-by-16 binning chosen to achieve a $3\sigma$ significance at least locally in the equivalent width image. The final images of Si and S were then smoothed, using the Smooth instrument in SAOImageDS9 to emphasize large-scale features.

\begin{figure}[t]
\begin{center}

\includegraphics[trim=0.0cm 0.0cm 0.0cm 0.0cm ,clip, scale=0.35]{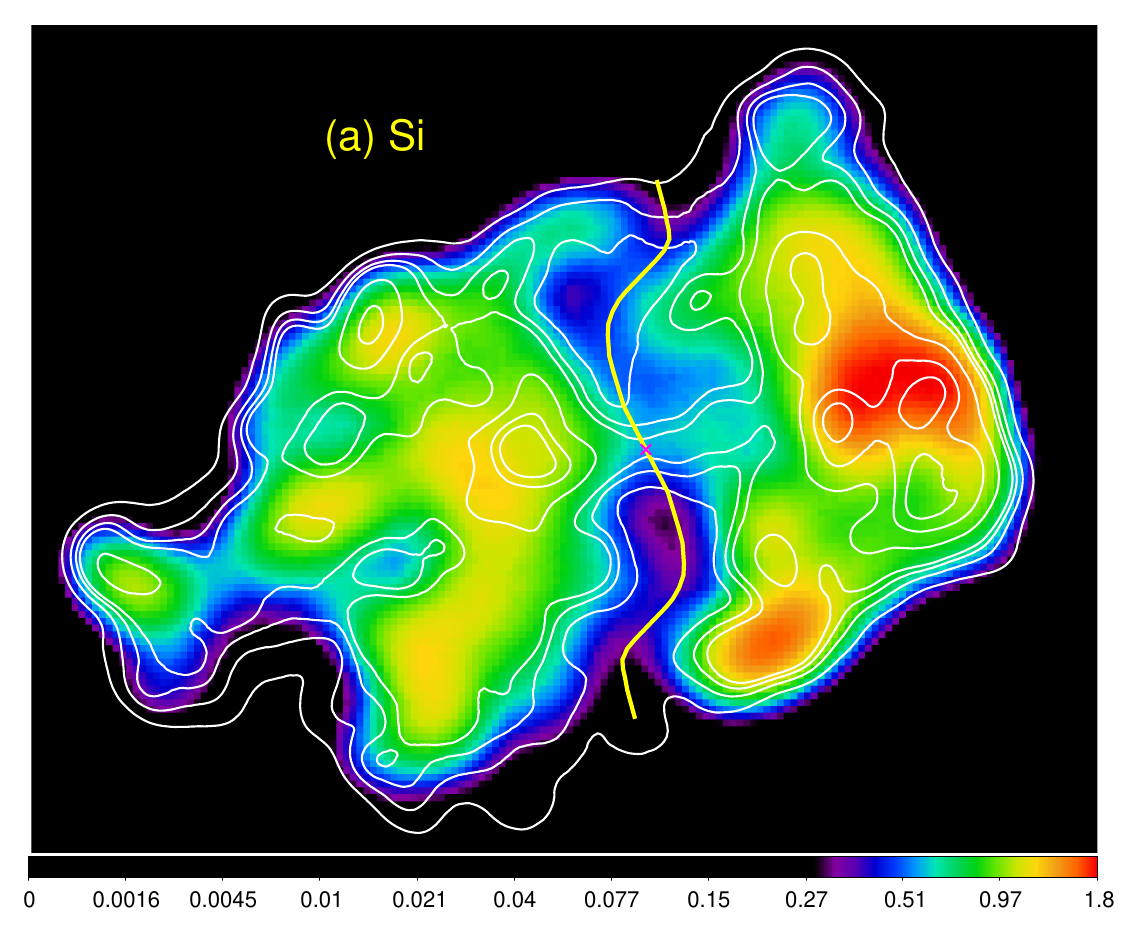}
\vskip 0.05 cm 
\includegraphics[trim=0.0cm 0.0cm 0.0cm 0.0cm ,clip, scale=0.35]{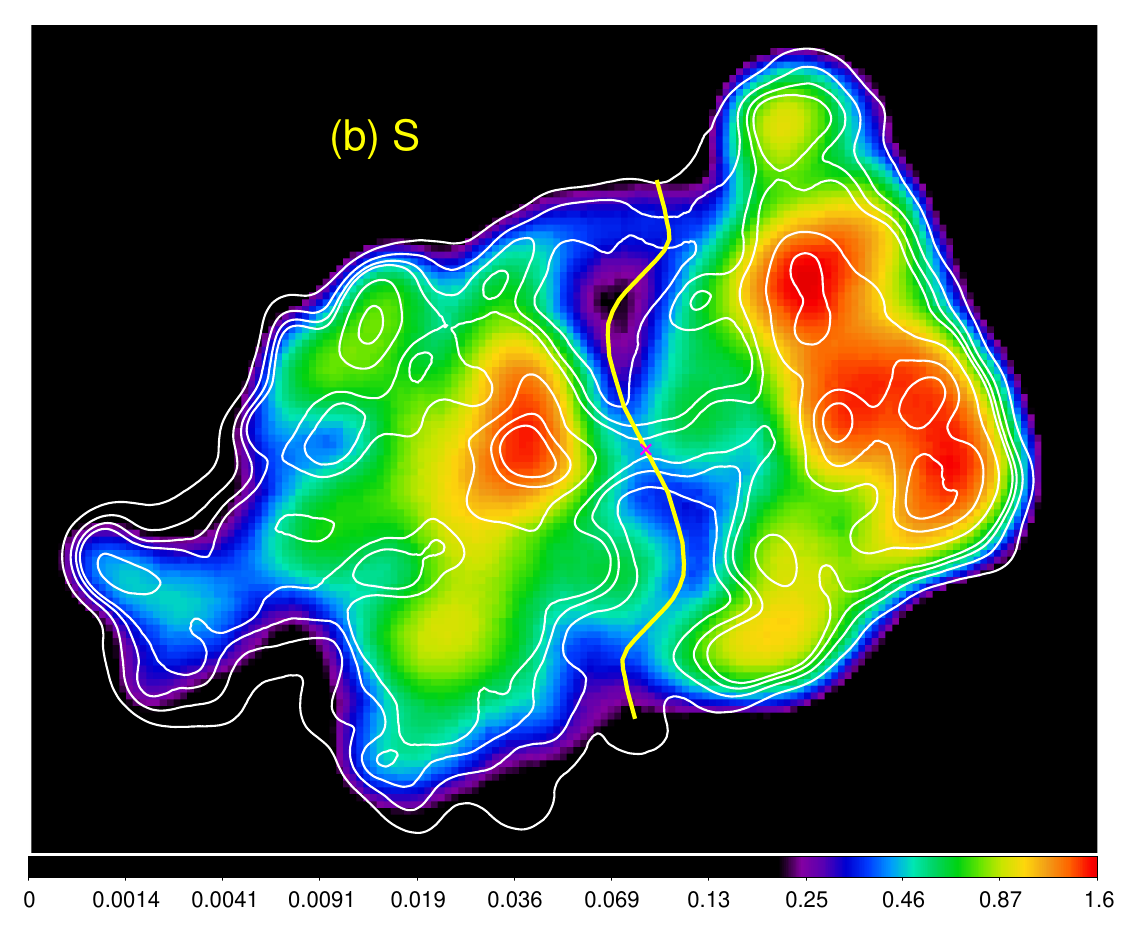}
\vskip 0.05 cm 
\includegraphics[trim=0.0cm 0.0cm 0.0cm 0.0cm ,clip, scale=0.35]{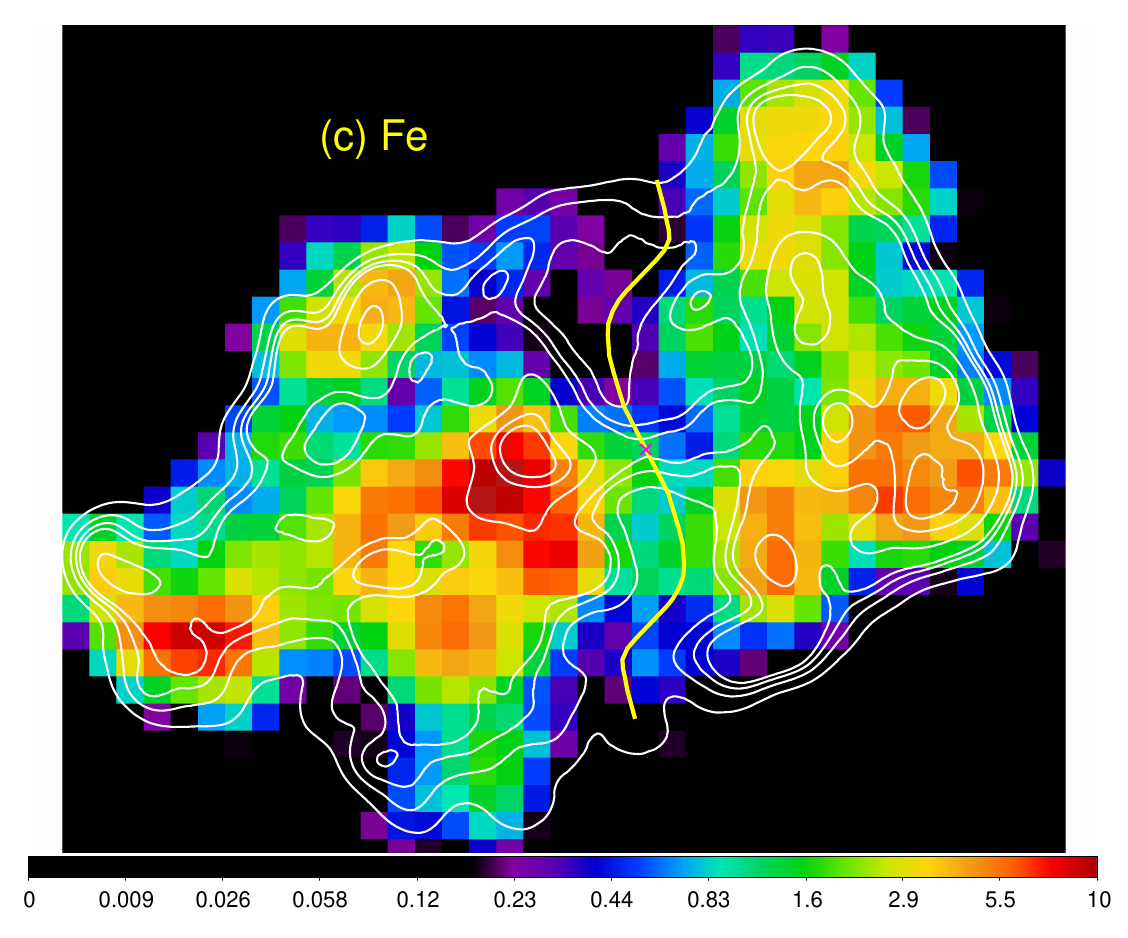}
\caption{X-ray maps of SNR 3C 397 in different lines as indicated in each panel. Contours and S-shaped yellow line as in Figure \ref{Fig:MaxSmLarge}.
Values of colors according to the different color bars. In panels (a) and (b), the images were binned with $4 \times 4$~pixels$^2$ and smoothed with a Gaussian function with $r=10~{\rm pixels}$ and $\sigma=5~{\rm pixels}$. In panel (c), the image was binned with $16 \times 16$~pixels $^2$.
The energy ranges in the equivalent width maps are as follows: (a) Si in the $1.78 - 1.9 \keV$ band,  
(b) S in the $2.35 - 2.5 \keV$ band, and (c) Fe in the $6.3 - 6.8 \keV$ band.  }
\label{Fig:lines}
\end{center}
\end{figure}

We present the equivalent-width images of three spectral lines in Figure \ref{Fig:lines}. The color maps show the equivalent width maps, while the S-shaped yellow line and the contours are as in Figures \ref{Fig:MaxSmLarge} and \ref{Fig:MediumSm}. The silicon and sulfur maps show a faint S-shaped zone extending from north to south through the center. The iron map has a large-scale structure compatible with the S-shaped faint region, but does not follow its detailed structure. We will see in Section \ref{sec:Perpendicular} that the iron map of SNR W49B, which has several similarities with SNR 3C 397, also does not follow the exact structure of the other elements.                     

Our main finding of this section is a robust, large, faint S-shaped structure in SNR 3C 397, which, based on hundreds of other astrophysical systems, including young stellar objects, planetary nebulae, X-ray binaries, and active galactic nucleus pairs of jets, we attribute to a pair of energetic jets. By energetic, we mean that the energy carried by the pair of jets is a substantial fraction of the explosion energy, possibly exceeding half.    

\section{Bright regions morphology}
\label{sec:Perpendicular}

\begin{figure*}[]
	\begin{center}
\includegraphics[trim=0.0cm 0.0cm 0.0cm 0.0cm ,clip, scale=0.7]{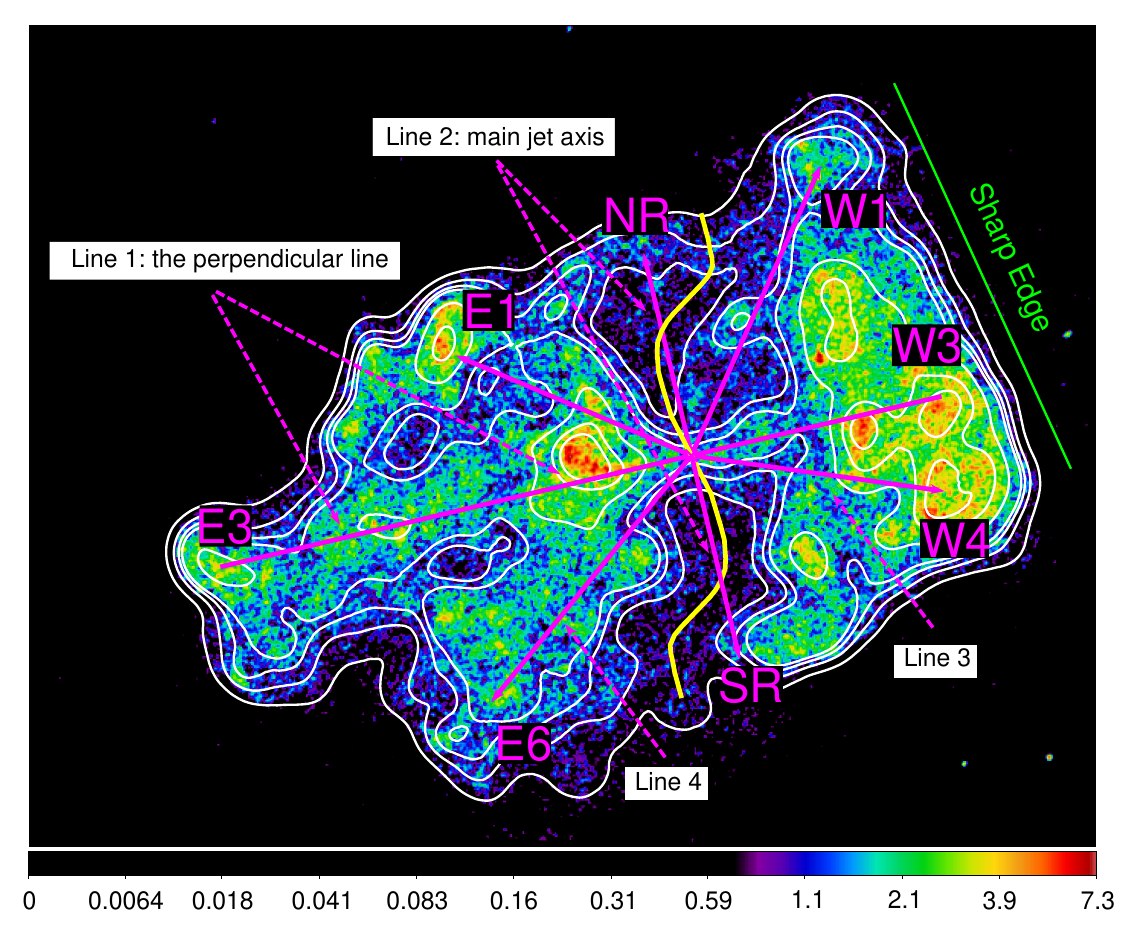} 
\caption{The X-ray image of 3C 397 from panel (b) of  Figure \ref{Fig:MediumSm}. We denote point symmetry by 4 lines connecting bright clumps on the west and east sides, and going through the center (see text). In lines L2, L3, and L4, the two opposite arrows are of the same length. In lines L1 and L2, which are perpendicular to each other, the two opposite sides are $180^\circ$ to each other; in lines L2 and L3, the angle between the two sides is $165^\circ$.}
\label{Fig:PerpStructure}
\end{center}
\end{figure*}

Following the identification of the central north-south S-shaped structure by the faintest regions of SNR 3C 397, we turn to examine the bright clumps on the west and east sides. In Figure \ref{Fig:PerpStructure}, we present an X-ray image as in panel (b) of Figure \ref{Fig:MediumSm}, and emphasize the following morphological features that relate bright clumps on the east and west sides. 
\begin{enumerate}
\item 
    \textit{Line L1: The perpendicular line.} We stretch a line along the longest dimension through the center, from clump E3 to clump W3. This line crosses several intensity peaks, E2, the south edge of E5, and the north edge of W5. We refer to this as the perpendicular line because it is perpendicular to the S-shaped structure.   
\item 
    \textit{Line L2: The main jet axis.} We draw a line perpendicular to L1 and through the center. It is a double-sided arrow to indicate that the center of this line coincides with the center of SNR 3C 397. We refer to it as the main jet axis of SNR 3C 397. In the scenario we propose, the pair of jets precesses around this axis (or an axis very close to it). 
\item 
    \textit{Line L3.} This is a broken line composed of two sides marked by arrows, one connects the center to clump E1 and one to clump W4. Both arrows are of the same length, and with an angle of $165^\circ$ between them. 
\item
    \textit{Line 4:} This is another broken line connecting clump W1 with clump E6. The two arrows are of the same length with an angle of $165^\circ$ between them.  
\end{enumerate}

Our claim is that the four lines of bright clumps and the S-shaped faint regions compose a point-symmetric morphology for SNR 3C 397. We do not argue that the lines L1, L3, and L4 are axes of pairs of jets. We can confidently identify only the S-shaped structure as a pair of precessing jets. Our point is that SNR 3C 397 exhibits a morphology that indicates some organized explosion, but with an asymmetrical component. The most asymmetrical component is the much larger eastward than westward extension of the SNR. Namely, the perpendicular line. 
This structure somewhat resembles the morphology of a similarly puzzling SNR, W49B.

\section{Comparison to W49B}
\label{sec:W49B}

SNR 3C 397 shares some properties and puzzles with SNR W49B. Most recently, \cite{Treyturiketal2026} discussed the similar (but not identical) abundance patterns of these two SNRs, with high iron abundances similar to those of SNe Ia but morphologies that better resemble those of CCSNe. We will return to this puzzle in Section \ref{sec:Summary} where we suggest two plausible scenarios for these SNRs. Here, we focus on some morphological similarities between the two SNRs.  

\begin{figure*}[]
	\begin{centering}
    \includegraphics[trim=0.0cm 0.0cm 0.0cm 0.0cm ,clip, width=\textwidth]{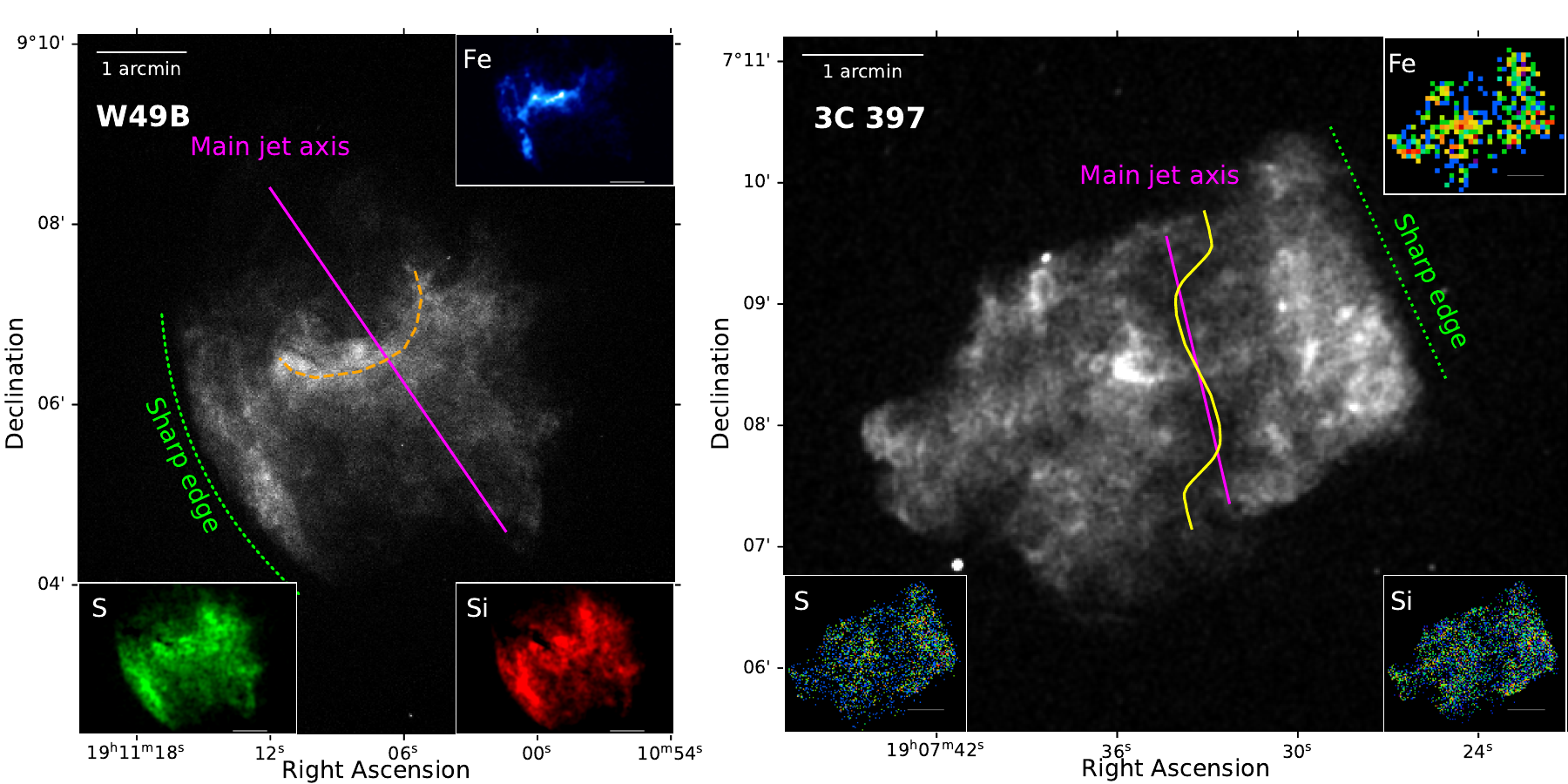} 
    \caption{Comparison of W49B (left, \citealt{Lopezetal2013a, SokerShishkin2025W49b}) and 3C 397 (right, Figures~\ref{Fig:MaxSmLarge}-\ref{Fig:PerpStructure} here and  \citealt{SafiHarbetal2005}). 
    \textbf{W49B}: Center panel: Total X-ray counts Chandra ACIS image, smoothed to $\sigma=1$. Magenta line denotes the main jet axis and the dashed orange ring the circum-jet ring \citep{SokerShishkin2025W49b}. The three panels at the corners of the image are smoothed line emission images of Fe (upper right), S (lower left), and Si (lower right), adapted from \cite{Lopezetal2013a}. 
    \textbf{3C 397}: Similar to the left panel of W49B but for 3C 397. Panels at the edges are as in Figure~\ref {Fig:lines}, except for Si and S being unsmoothed. Note that the color scales differ from panel to panel to highlight the distribution rather than intensity. The yellow curve is the S-shape from Figure~\ref{Fig:MaxSmLarge}. A white horizontal line in all panels is a $1'$ scale bar. \\
    In both W49B and 3C 397, the iron emission distribution is noticeably different from that of the total and other lines. Both remnants feature a similar fainter central region in the emission, which we attribute to a main jet axis (magenta line); see \cite{SokerShishkin2025W49b} for axis definition. Both remnants also exhibit a single sharp edge (green line; W49B: left; 3C 397: right). }
    \label{Fig:W49B3C397}
    \end{centering}
\end{figure*}

In Figure \ref{Fig:W49B3C397} we present images of the two SNRs. 
We find the following common properties in the two SNRs' morphologies. 
\begin{enumerate}
\item 
    Both SNRs exhibit a clear main jet axis. The one in W49B is a straight line of two opposite jets (e.g., \citealt{BearSoker2017}), identified by an ear in the southwest, a circum-jet ring near the center, and a clump in the northeast (\citealt{SokerShishkin2025W49b}). In 3C 397, the main jet axis is of a precessing pair of jets. 
\item 
    The two sides, more or less perpendicular to the main jet axis, are not symmetrical. In both SNRs, the eastern side extends farther than the western side from the main jet axis. 
\item 
    One side is brighter and has a sharp, almost straight edge: in W49B, it is the eastern side; in 3C 397, it is the western side. 
\item  
    In both SNRs, there is a thin bar connecting the two sides. In W49B, it is a ring structure (dashed orange ring in Figure~\ref{Fig:W49B3C397}; \citealt{SokerShishkin2025W49b}).  
\item 
    In both SNRs, the iron morphology differs from the general morphology of the total intensity and from the morphologies of other elements. 
\end{enumerate}

We point out that the main jet axis we adopt for W49B, which is based on \cite{BearSoker2017} and \cite{SokerShishkin2025W49b}, is not in consensus. \cite{XRISMW49B2025} analyzed the velocity structure of W49B and argued for a symmetric axis of a bipolar explosion that is the same as some earlier claims (e.g., \citealt{Keohaneetal2007, Lopezetal2013a}), and which is more or less perpendicular to the direction we take for the main jet axis.  In the models we discuss in Section \ref{sec:Summary}, we attribute the velocity structure that \cite{XRISMW49B2025} analyzed to an asymmetrical mass ejection in the equatorial plane, as evident from the two unequal sides of W49B, as well as of 3C 397. 

The similarity in many properties of W49B and 3C 397 implies that these types of SNRs are, although rare, not extremely rare. Any scenario for their formation should incorporate this implication, as we further discuss in Section \ref{sec:Summary}.  

\section{Discussion and Summary}
\label{sec:Summary}

We analyzed X-ray observations of the enigmatic SNR 3C 397, and identified a faint north-south region with a clear S-shaped morphology (Figure \ref{Fig:MaxSmLarge}-\ref{Fig:lines}). We attributed the shaping of this structure to a precessing pair of jets that carry a significant fraction of the explosion energy, possibly more than half, as evident by the large volume of the jet-inflated bubbles.   
In Section \ref{sec:Perpendicular} we discuss the east-west extension of a chain of bright knots, which we term the equatorial plane and mark with line L1, the perpendicular line, in Figure \ref{Fig:PerpStructure}. We suggested there that the pair of jets precess around an axis, termed L2, perpendicular to this plane and passing through the center of the SNR, which we identified as the point marked by a red X in Figure \ref{Fig:MaxSmLarge}-\ref{Fig:lines}; this axis (L2) is the main jet axis of SNR 3C 397. We also noted two other axes, with $165^\circ$ bent around the center (lines L3 and L4). The lines L1-L4 compose the point-symmetric structure of SNR 3C 397, which, together with the S-shaped bubbles, robustly point at shaping by jets. We identify one pair of jets, but cannot rule out more pairs of weaker jets that contributed to the explosion.  

In Section \ref{sec:W49B}, we pointed out some morphological similarities of SNR 3C 397 with SNR W49B, particularly a main jet-axis for an energetic pair of jets, unequal sides, a thin bright bar that connects the two sides, and the different morphology of iron with respect to some other metals. 

Several studies have noted the puzzling nature of these two SNRs (e.g., \citealt{Treyturiketal2026} and some references therein). We mentioned some puzzles concerning SNR 3C 397 in Section \ref{sec:Introduction}. A pronounced disagreement concerns the progenitor of SNR W49B, with some studies arguing for a thermonuclear explosion as an SN Ia (e.g., \citealt{Hwangetal2000, ZhouVink2018, Siegeletal2020, Satoetal2024}), some for a CCSN (e.g., \citealt{Lopezetal2011, Lopezetal2013a, Yamaguchietal2014, Patnaudeetal2015}), and one study for a CEJSN with thermonuclear outburst \citep{GrichenerSoker2023}. In addition, some studies find that W49B does not fit either the CCSN or the SN Ia scenarios (e.g., \citealt{Patnaudeetal2015, Siegeletal2020}). \citet{Sawadaetal2025} argue that the Fe-group ejecta mass ratios might result from either an SN Ia or a CCSN. 

In discussing scenarios to explain SNR 3C 397 and W49B, we consider the following results. 
($i$) Our finding of an S-shaped morphology in SNR 3C 397 strongly suggests that a precessing pair of jets shaped a large fraction of the volume of SNR 3C 397. The precession is around an axis, the major jet axis of 3C 397. 
($ii$) The claim of a major jet axis in W49B \citep{BearSoker2017}, which later studies adopted (\citealt{Akashietal2018, Siegeletal2020, GrichenerSoker2023}), and which \cite{SokerShishkin2025W49b} solidified. 
($iii$) The recent study by \cite{Treyturiketal2026} found some similar composition puzzles in SNRs W49B and 3C 397. The composition and explosion energy of SNR 3C 397 suggest some kind of thermonuclear explosion, but not that of a regular SN Ia. 

We mention two possible theoretical scenarios for both SNR 3C 397 and W49B, both of which need much more exploration. 

\textit{(1) The thermonuclear CEJSN scenario.} 
In the thermonuclear CEJSN scenario first suggested by \cite{GrichenerSoker2023} for SNR W49B, the NS tidally disrupts the core of a red supergiant star, forming an accretion disk around the NS. Such tidal disruption can trigger a thermonuclear outburst in the accretion disc  (e.g. \citealt{Fryeretal1999, Zenatietal2019}; also, \citealt{YeCetal2023} who discussed the tidal disruption of a WD by an intermediate mass black hole, resulting in jets and thermonuclear outburst). \cite{Zenatietal2020} simulate the disruption of a WD by an NS. For a WD mass of $M_{\rm WD} = 0.8 M_\odot$, the ejected mass of iron group elements is $M_{\rm IGE} = 0.045 M_\odot$. \cite{Bobricketal2022} simulated the merger of an NS with an ONe WD. They found that the mass of $^{56} {\rm Ni}$ that the event produces growth as $M_{\rm 56Ni} \approx 0.04 (M_{\rm WD}/M_\odot)^{3.4} M_\odot$. Substituting a core mass of $2 M_\odot$, yields $M_{\rm 56Ni} \approx 0.4 M_\odot$. 
This might explain the iron group elements mass of $M_{\rm IGE,3C397} \simeq 0.32 M_\odot$ that \cite{Treyturiketal2026} estimated for 3C 397. As the thermonuclear CEJSN scenario requires the NS to tidally disrupt the core, the core mass cannot be too large; \cite{GrichenerSoker2023} estimated the core mass for this scenario to be $2 M_\odot \lesssim M_{\rm core} \lesssim 3.5 M_\odot$. The same thermonuclear CEJSN discussed by \cite{GrichenerSoker2023} might also account for SNR 3C 397. 
The process of tidal disruption ejects mass in a non-axially symmetric morphology, and might account for the unequal sides of such SNRs. 
\cite{GrichenerSoker2023} estimated that there is one CEJSN where the NS tidally destroys the core for 200 CCSNe. This is a very low rate, and that of cases with a thermonuclear explosion accompanying the launching of jets is lower still. Having two such SNRs among fewer than 100 well-resolved SNRs with still intact morphology is somewhat larger than expected. This deserves further study. 

\textit{ (2) The collapse-induced thermonuclear jet-driven explosion.} 
In the collapse-induced thermonuclear explosion mechanism, a mixed layer of helium and oxygen undergoes thermonuclear burning as it collapses and heats up, and this is supposed to cause the star to explode (\citealt{Burbidgeetal1957, KushnirKatz2015}). For that to occur, the pre-collapse core must have a large amount of angular momentum (e.g., \citealt{Kushnir2015b, Gofmanetal2018}). As a result of that, a very massive accretion disk is formed around the newly born neutron star (\citealt{Gilkisetal2016}; later confirmed by \citealt{BlumKushnir2016}). The expectation is that the energy carried by the jets exceeds the energy released by thermonuclear reactions (e.g., \citealt{Gilkisetal2016}). Therefore, we term this scenario a collapse-induced thermonuclear jet-driven explosion.   
We here raise the possibility that the progenitors of SNRs W49B and 3C 397 had rapidly rotating pre-collapse cores and helium-oxygen mixed layers. These rare conditions formed a long-lived, fixed-axis accretion disk that launched very energetic jets, and the mixed layer triggered an energetic thermonuclear outburst.  
Future studies should explore the quantitative parameters of this scenario and whether it can occur at all. 
The unequal sides of SNRs W49B and 3C 397, in this case, suggest that the thermonuclear outburst of the helium-oxygen mixed layer was asymmetrical, e.g., starting at one point and propagating around the explosion center.  
 \citet{Satoetal2024} claim that their determined titanium abundance in SNR W49B excludes almost all hypernova/jet-driven supernova models. However, they did not consider the thermonuclear outburst induced by collapse. 
In a recent study, \cite{Kozyrevaetal2026} proposed a scenario for the Helium superluminous SN 2021bnw, where the explosion is jet-driven, and forms about $1.7 M_\odot$ of $^{56}$Ni. This case might well be a thermonuclear jet-driven supernova induced by collapse.  
We emphasize here that the jet-driven mechanism we discuss is part of the JJEM \citep{Soker2026G11}; it is one extreme edge where the pre-collapse core rotation contains much more angular momentum than the fluctuating angular momentum components due to the convection in the pre-collapse core (the other edge is a non-rotating core). 

\cite{Treyturiketal2026} apply the Sedov blast wave model under a few assumptions, and estimated the explosion energy of the two SNRs to be $\simeq 10^{50} \erg$. In both scenarios that we propose, we expect a much larger explosion energy, i.e., $\gtrsim 10^{51} \erg$.

Future theoretical studies based on existing and new observations should determine whether either of these two scenarios, or a different one, can account for SNRs 3C 397 and W49B. None of these scenarios can ignore the robust evidence for an energetic pair of jets in each SNR. 



\software{
CIAO \citep{Fruscione2006},
SAOImage DS9 \citep{William2017},
Astropy \citep{astropy:2013, astropy:2018, astropy:2022},
Matplotlib \citep{Hunter:2007},
          }

\paragraph{Funding Statement}
A grant from the Pazy Foundation 2026 supported this study. 
NS thanks the Charles Wolfson Academic Chair at the Technion for the support.

\bibliography{bib.bib}{}
\bibliographystyle{aasjournal}

\end{document}